%% file: main.tex
\documentclass{elsart}
\pdfoutput=1

\usepackage{graphicx}
\usepackage{color}
\usepackage{amsmath}
\usepackage{tabularx}
\usepackage{supertabular}
\usepackage{longtable}
\usepackage{booktabs}
\usepackage{float}
\usepackage[hyphens]{url}

\begin{document}

\begin{frontmatter}

\title{The Eclipse Integrated Computational Environment}

\author{Jay Jay Billings$^{1,2}$}
\ead{billingsjj@ornl.gov, Twitter: @jayjaybillings}
\author{Andrew R. Bennett$^{1,3}$}
\author{Jordan Deyton$^{1,4}$}
\author{Kasper Gammeltoft$^{1,5}$}
\author{Jonah Graham$^{6}$}
\author{Dasha Gorin$^{1,7}$}
\author{Hari Krishnan$^{8}$}
\author{Menghan Li$^{9}$}
\author{Alexander J. McCaskey$^{1}$}
\author{Taylor Patterson$^{1,10}$}
\author{Robert Smith$^{1}$}
\author{Gregory R. Watson$^{1}$}
\author{Anna Wojtowicz$^{1,11}$}

\address{$^{1}$Computer Science and Mathematics Division, Oak Ridge National
Laboratory, Oak Ridge, TN 37830, USA}
\address{$^{2}$The Bredesen Center for Interdisciplinary Research and Graduate
Education, University of Tennessee, 444 Greve Hall, 821 Volunteer
  Blvd. Knoxville, TN 37996-3394}
\address{$^{3}$University of Washington, Seattle, WA 98105}
\address{$^{4}$General Electric Company, 3200 North Grandview Blvd Waukesha, WI
53188-1678}
\address{$^{5}$Georgia Institute of Technology North Avenue, Atlanta, GA 30332}
\address{$^{6}$Kichwa Coders Ltd. 1 Plomer Green Avenue Downley, High Wycombe
HP13 5LN United Kingdom}
\address{$^{7}$Northwestern University 633 Clark Street Evanston, IL 60208}
\address{$^{8}$Lawrence Berkeley National Laboratory, 1 Cyclotron Rd, Berkeley,
CA 94720}
\address{$^{9}$Department of Computer Science, and Department of Biological
Sciences, Purdue University, West Lafayette, IN 47906} 
\address{$^{10}$Acato Information Management, LLC 114 Union Valley Rd. Oak Ridge,
TN 37830}
\address{$^{11}$Colorado State University Fort Collins, CO 80523}

\begin{abstract}

Problems in modeling and simulation require significantly different
workflow management technologies than standard grid-based workflow
management systems. Computational scientists typically interact with
simulation software in a feedback-driven way where solutions and
workflows are developed iteratively and simultaneously. This work
describes common activities in workflows and how combinations of these
activities form unique workflows. It presents the Eclipse Integrated
Computational Environment as a workflow management system and
development environment for the modeling and simulation community.
Examples of the Environment's applicability to problems in energy
science, general multiphysics simulations, quantum computing, and other
areas are presented along with its impact on the community at large.

\end{abstract}

\begin{keyword}
workflows
\sep 
workflow management
\sep
supercomputing
\sep
usability
\sep
eclipse

\vspace{1ex}

\PACS
07.05.Tp 

\end{keyword}

\end{frontmatter}

\section{Notice of Copyright}\label{notice-of-copyright}

This manuscript has been authored by UT-Battelle, LLC under Contract No.
DEAC05-00OR22725 with the U.S. Department of Energy. The United States
Government retains and the publisher, by accepting the article for
publication, acknowledges that the United States Government retains a
nonexclusive, paid-up, irrevocable, world-wide license to publish or
reproduce the published form of this manuscript, or allow others to do
so, for United States Government purposes. The Department of Energy will
provide public access to these results of federally sponsored research
in accordance with the DOE Public Access Plan
(http://energy.gov/downloads/doe-public-access-plan ).

\input{content}

\bibliographystyle{unsrt}
\bibliography{bib}

\end{document}

%% file: content.tex
\section{Motivation and Significance}\label{motivation-and-significance}

In previous work, Billings et al. interviewed modeling and simulation subject matter experts to compile a list of requirements for implementing and 
using these kinds of applications. In the process, they discovered 
that many of the difficulties inherent in using
high-performance modeling and simulation software fall into five
distinct categories \cite{billings_designing_2009}. These activities,
detailed in Section \ref{workflow-model}, include (1) creating input,
(2) executing jobs, (3) analyzing results, (4) managing data, and (5)
modifying code. There are many tools that address these problems
individually, but the same research found that the excess number and
specialization of these tools also contribute to the learning curve.

Previous efforts to address these five issues have resulted in general
purpose scientific workflow tools like Kepler \cite{ludascher_scientific_2006} 
or myopic tools that only satisfy a single
set of requirements for a single piece of software or a single platform. These
are opposite extremes, but a middle-of-the-road solution is also
possible. A workflow engine could be developed that limits its scope to
high-performance computing (HPC) and to the set of possible workflows
associated with the five previously mentioned activities. With only
minor additional development, a rich application programming interface
(API) could be exposed so that highly customized solutions could still
be made based on this limited workflow engine.

It is not clear which, if any, of these solutions is better than the
others, and practical requirements will ultimately dictate the path of a
project's progress. This work considers a middle ground solution and
presents the Eclipse Integrated Computational Environment (ICE) as proof
that it is possible to create such a system. Specifically, the work
described here shows that:

\begin{itemize}
\item
  Modeling and simulation activities can be described in a succinct
  workflow model (see ``Workflow Model'').
\item
  An architecture for such a workflow system can satisfy the model of
  workflows in an extensible way (see ``Software Architecture'').
\item
  Such a system is applicable to a suite of problems in energy science,
  including virtual battery simulations and additive manufacturing, 
  among others (see ``Illustrative Examples'').
\end{itemize}

This section concludes with an introduction of the ICE workflow model. Section \ref{software-description} details the software
from an architectural perspective, while Section \ref{illustrative-examples}
provides a set of comprehensive examples. A
presentation of the impact is included in Section \ref{impact}, and sample 
code is provided in Section \ref{sample-code-tutorials-and-other-resources}.

\subsubsection{Workflow Model}\label{workflow-model}

Computational scientists perform a variety of tasks in modeling and
simulation that can be abstracted into a lightweight theoretical
framework based broadly on five high-level activities: (1)
creating input, (2) executing jobs, (3) analyzing results, (4) managing
data, and (5) modifying code. Those same computational scientists would
most likely find these activities difficult for all codes with which
they lack experience, whereas with their own codes---or those with which
they are most familiar---these tasks may be so simple that they are
taken for granted. Any particular combination of these activities across
one or more scientific software package or code results in a unique
workflow. Such a workflow is normally, but not always, requested
by a human user and orchestrated by a workflow management system.

The most obvious workflow for any individual simulation code or
collection of codes is to string the activities together, where the user's
workflow is to create the input, launch the job, perform some analysis,
and manage the data---possibly modifying the code in the process. However, 
there are many other combinations, including re-running jobs with
conditions or modifications or analyzing someone else's data.\footnote{The
authors have identified many unique combinations that
define workflow ``classes.'' When possible, every effort is made to give the
classes colloquial names such as ``The Re-Run'' or ``The Graduate Student.''}

\textbf{Creating input} is the process of describing the physical model
or state of a system that will be simulated. This could include creating
an input file(s) or making calls to an external process to configure a
running program. In most situations, a computational scientist will
modify existing input or create new input from a template. ``Input''
generally includes runtime parameters for the simulation framework
(e.g., tolerances); configuration options (e.g., data locations, output
locations, module configurations); properties of the materials to be
simulated; and a discretization of the simulation space (e.g., mesh,
grid, particle distribution). The collection of all required input can
be quite large and may go by many names, including ``input set,''
``input package,'' ``problem,'' or, simply, ``input.'' Often, the set of
input files will be described in a ``main'' input file that acts as a
kind of manifest to describe---and provide links to---all necessary
information for a given problem.

In this work, it should be assumed---unless otherwise noted---that
``input'' refers to the entire set of input, not to a single file.

\textbf{Executing jobs}, or ``running the workflow'' in this context, is
the process of performing calculations using a simulation code or
framework based on known variables---the input. Models and simulation
codes are typically run locally for small jobs or for development. Large
simulations, on the other hand, typically require a large amount of
hardware resources; these resources are usually off site (i.e.,
physically unavailable to the user) and are accessed remotely
through Secure Shell (SSH) connections or similar protocols. Remote
execution requires moving the input in advance of the execution and
copying or moving the output to the user's machine. In many cases,
though, the output is too large to move to the user's local machine.

Local and remote jobs are often monitored to ascertain a job's status.
This monitoring may be a simple check as to whether or not the execution
has completed, or it may involve monitoring the output of individual
quantities to examine the calculation state. The latter is often used to
detect calculation errors that will result in incorrect results. If such
problems are found, the job is typically cancelled (``killed'') to save
compute resources and is then re-run later.

In this work, it should be assumed---unless otherwise noted---that
``executing a job'' includes monitoring that job in one or more ways,
possibly including real-time updates to visualizations.

\textbf{Analyzing results} includes executing special jobs to transform
data in one or more prescribed ways and producing artifacts with
scientific significance from the transformed data. This may include, for
example, post-processing results and visualizing the new data with
dedicated visualization tools. For many types of scientific computing,
this includes viewing the results of a simulation on a mesh or grid and
extracting publication-quality images or movies from that data. Other
cases may include analyzing results in preparation for follow-on
simulations or performing feature extraction, classification, or
activities for machine learning and data mining.

While this has many similarities to executing a job, it is distinctly
different because the activity changes focus to satisfy the needs of a
human operator. Simple data reduction, where the exact reduction is
known, certainly qualifies as executing a job; however, analysis of
model and simulation results is far from simple data reduction and is
generally far more interactive for computational scientists.

\textbf{Managing data} includes moving, copying, storing, sharing, or
otherwise interacting with data for or from simulations. This activity
is the most pervasive because each of the other activities requires
interacting with data in some way. In many cases, though, data is still
managed for its own purposes, without performing a simulation,
generating new input, or analyzing results. Examples include archiving
data, packaging data for publications, and updating values manually
(often in light of new information from publications).

\textbf{Modifying code} is not typically considered a part of a
scientific computing workflow. However, modeling and simulation use
cases often require users to explicitly modify code before execution,
such as with the computational fluid dynamics code Nek5000,
\cite{the_nek5000_team_nek5000_2014}, or to issue special re-build instructions.
Many computational scientists consider ``their workflow'' to be re-running
software after modifications for purely exploratory purposes. This may
be required if the model that the author is manipulating cannot be
configured directly as part of the input but can be easily manipulated
by hand.

\subsubsection{Comparison to Other
Models}\label{comparison-to-other-models}

This model of workflows differs significantly from those of similar
efforts in workflow science because it defines workflows in terms of
activities. Other workflow models in the literature define a workflow as
a collection of computing processes. For example, Yu and Buyya define
grid workflows as ``a collection of tasks that are processed on
distributed resources in a well-defined order to accomplish a specific
goal,'' \cite{yu_taxonomy_2005}. Others, such as Pizzi et al., subscribe to
similar definitions \cite{pizzi_aiida:_2016}. This ``process'' view is
acceptable where the workflow is static and does not require additional
human input or ``human in the loop'' behavior after all the initial
human input is provided. That is, a ``process-oriented'' definition is
acceptable where all human input is provided in advance. However,
workflows within ICE are fully interactive with regular call-backs to
humans. It is simpler to discuss ``activities'' than it is to create a
distinction between ``human processes'' and ``computer processes.''
Focusing on ``activities'' over processes (human or computer) also has
the benefit of removing concrete elements such as hardware properties or
software details that distract from details of workflows and workflow 
management systems. That is, considerations such as memory usage and 
raw performance are important, but questions about the abstract workflow 
or what the workflow management system should do are far more important.

\section{Software Description}\label{software-description}

ICE was specifically created to address the workflow model described
above, which is to say that it was created specifically to address
hands-on workflows for computational scientists. Users download and
execute ICE locally and it orchestrates workflows locally or remotely as
required. It provides a comprehensive workbench for modeling and
simulation that includes tools for workflows, visualization, data
management, and software development.

\subsection{Software Architecture}\label{software-architecture}

Workflows and tasks in ICE are not explicitly treated as trees, or
directed acyclic graphs (DAGS), as is common with grid workflow tools
\cite{yu_taxonomy_2005}. Instead, ICE's design is inspired by representational
state transfer (REST) \cite{fielding_architectural_2000}.

Users initiate requests to create, edit, update, or delete workflows from
the ``ICE Client'' (the workbench). The list of available workflows
that can be created is provided dynamically to the ICE Client by
the ``ICE Core,'' which acts as a server. This information is
provided dynamically because the information often changes at run time based on 
both the configuration of available workflow components in the registry and on
persisted workflows that users have saved in their workspace.
Information about workflows is provided to the Client by the Core
through stateless ``Forms'' that describe the workflow and provide
all the necessary information to understand \emph{what} should process
the workflow (but not \emph{how} it should be processed). Once users
modify the description of the workflow in the Form to provide their
specific details, the Client dispatches a request to the Core to modify
or process the workflow. The Core then uses the information from the Form
to perform a service lookup for \emph{what} should process the workflow.

Workflows in ICE are encoded in and processed by ``Items'' and each
workflow type is a subclass of Item. Items are registered dynamically
through a service registry in ICE (see Section \ref{framework}) and
provide the Forms for the ICE Core and Client. Since ICE's design is
highly object oriented, it is easiest to think of the Item class as a
description of an abstract workflow and an Item object (an instance of
the class) as a concrete workflow with all required execution details
provided by its Form.

Individual components of workflows (i.e., workflow ``tasks'' or
``nodes'') are either encoded directly in the workflow's subclass of
Item or provided as ``Actions'' that are dynamically registered with
an ``Action Factory'' and obtained at runtime. Table 1 describes the
differences between Items, Actions, and Forms.\footnote{An upcoming 
update to the API will include the formal introduction of
IWorkflow, IWorkflowTask, and IWorkflowEngine interfaces to bring ICE's
API language closer to other systems. This must be done carefully to
preserve backwards compatibility with the present API.}

\begin{table*}[t]
\begin{tabularx}{\textwidth}{|l|X|l|}
\hline
Class & Class Description & Object Description\tabularnewline\hline
Item & Java class with code to execute an abstract workflow. Provides a
Form. & Concrete workflow executor.\tabularnewline\hline
Form & Description and template of the data needed for the Item to
process the workflow. & User-modified workflow data.\tabularnewline\hline
Action & Java class for executing a specific task in the workflow. Used
by Items. & Concrete workflow task executor.\tabularnewline\hline
\end{tabularx}
\caption{Class Descriptions for Items, Forms, and Actions}
\end{table*}

In addition to running as a desktop workbench, ICE can be run as a
headless web server with a remote service interface and a web
API. The web API is also used as the primary means of providing real-time 
feedback and monitoring support in ICE.

\subsubsection{Item States}\label{item-states}

All Items in ICE are finite state machines where the states represent
the abstract state of the workflow. For example, when an Item is first
created, it enters the ``Form Ready'' state to indicate that it could,
in theory, be processed after a user reviews it. After that review, it
enters the ``Ready to Process'' state before it is processed and 
the ``Processed'' state after it is processed. There are several additional 
states for errors.

This design is very important. First, it means that all workflows in
ICE, regardless of their actual details and functions, can only behave in
a specific set of known and predictable ways, and this predictability 
simplifies the way the Core manages and interacts with Items. Second, by
formalizing state and error checks, ICE explicitly delineates which workflows 
can be executed from those that must receive additional configuration. 
Finally, it allows developers implementing Items and
Actions to specify, by contract, what is required before proceeding to the
next task, processing the workflow, or declaring a successful execution.

\subsubsection{Persistence and
Workspaces}\label{persistence-and-workspaces}

When workflows are created and modified, ICE saves permanent copies of 
Forms to disk in a special directory called a ``workspace.''
Workspaces can contain projects, folders, and files; including data,
code, input, and output. ICE automatically manages local and remote (or
even local \emph{to} remote) transfers of data files when executing
workflows if the files are detected in the same directory of the
workspace as the workflow itself. For example, when executing a remote
job, ICE will automatically move the input file if it is specified in
the workflow and available in the project directory of the workspace.
Likewise, if the output is small enough, ICE will automatically move it back
to local directory. By convention, all paths in ICE's Forms are
relative to the workspace root path. Workspace directories are specified
by the user.

ICE handles persistence using a ``persistence provider.'' The
default persistence provider uses JAX-RS to write Forms to 
XML~\cite{burke_restful_2010}. In principle, other persistence providers 
could replace the XML-based provider since it is registered as a dynamic 
service, and a JAX-P based provider has been used in the past.

\subsubsection{Extending the Framework to Add Custom
Workflows}\label{framework}

ICE is an Eclipse Rich Client Platform (RCP) application
\cite{mcaffer_eclipse_2010}, and has a plugin architecture based on 
Equinox, which is the reference implementation of the Open Service Gateway Initiative (OSGi) 
framework specification \cite{mcaffer_osgi_2010}. ICE uses over 1,200 additional
packages from the Eclipse ecosystem to provide services like language
support and visualization. Each unique element of ICE described
in this work---including the Client, Core, Items, and data structures---are 
provided as plugins to Equinox. Most plugins are managed
dynamically and provided as services that can be obtained through a
service registry. Most file I/O in ICE, with only a few exceptions,
interacts with the RCP's resources plugin. This also includes remote
resources that are managed with the Eclipse Parallel Tools Platform
\cite{tibbitts_integrated_2009}.

It is not possible to create new Items in ICE graphically, and---unless 
they provide their own plugin to the framework---users can
only use the workflow Items provided by their version of ICE. This can seem like an
onerous task to the novice user, especially if they do not know Java, but
this design was chosen because it is, in the opinion of the authors, a
far more sophisticated and sustainable way to create workflows than, for example,
using a wiring diagram.

ICE provides built-in development tools to automatically generate ``stubs'' of
plugins that can then be installed into the framework. This
``self-hosting'' makes it possible for new users to create complex,
sophisticated workflows very quickly because they are not required to
know how to interact with the framework. Furthermore, ICE provides a
rich API, documentation, tutorials, and tools to further simplify the
process.

There are many situations where configuring workflows graphically is
unacceptable, such as when a very large number of workflows will be
executed or the type of information required is very fine-grained. In
these cases, it is often necessary to provide scripts to the workflow
management system. ICE includes the Eclipse Advanced Scripting
Environment (EASE)~\cite{pontesegger_eclipse_2015} for scripting because it
provides a way to script Eclipse RCP projects natively in Javascript,
Jython, and Python. This also makes it possible to extend the
environment by adding Items in these languages.

\subsection{Software Functionalities}\label{software-functionalities}

The most important function of ICE is to serve as an easily extended
workflow management system for computational scientists in support of
the activities described above. In practice, it is most often used as a
combination of workflow management system and development environment
since it contains a significant amount of Eclipse's software development
tooling in addition to the workflow tools. The project has many users,
but it is most heavily used by the development team to support workflow
science and software development for energy science projects. The
development team regularly uses the platform to quickly deploy new
domain-specific workbenches in a matter of hours for small collections
of workflows that are easy to encode.

Outside of the development team, ICE is commonly deployed as a
sophisticated user environment for computational science projects 
(see Section \ref{impact}) and as a visualization tool. The ICE source code originally contained a
significant amount of visualization support, but at the request of users
in the community that support was ``spun off'' in early 2016 as the 
Eclipse Advanced Visualization Project (EAVP)~\cite{billings_eclipse_2015}.

\section{Illustrative Examples}\label{illustrative-examples}

The role that ICE plays as a workflow tool is best illustrated by the
various ways in which it has been deployed, as shown in the following 
examples.

\subsection{Virtual Battery
Simulations}\label{virtual-battery-simulations}

Pannala et al. developed a Virtual Integrated Battery Environment (VIBE)
as part of their research into safety and performance characteristics of
lithium ion batteries \cite{pannala_multiscale_2015}. VIBE includes ICE
as part of its distribution, and new workflows were added to ICE to enable users
to add multiple types of input, configure the simulation software, and launch
simulations of virtual batteries. Interactive 3-D visualizations of the
results were embedded in the launcher so that users can quickly find
their results. Figure 1 shows a (simulated) prismatic-cell battery's temperature
distribution during discharge.

\begin{figure}[htbp]
\centering
\includegraphics[width=\textwidth]{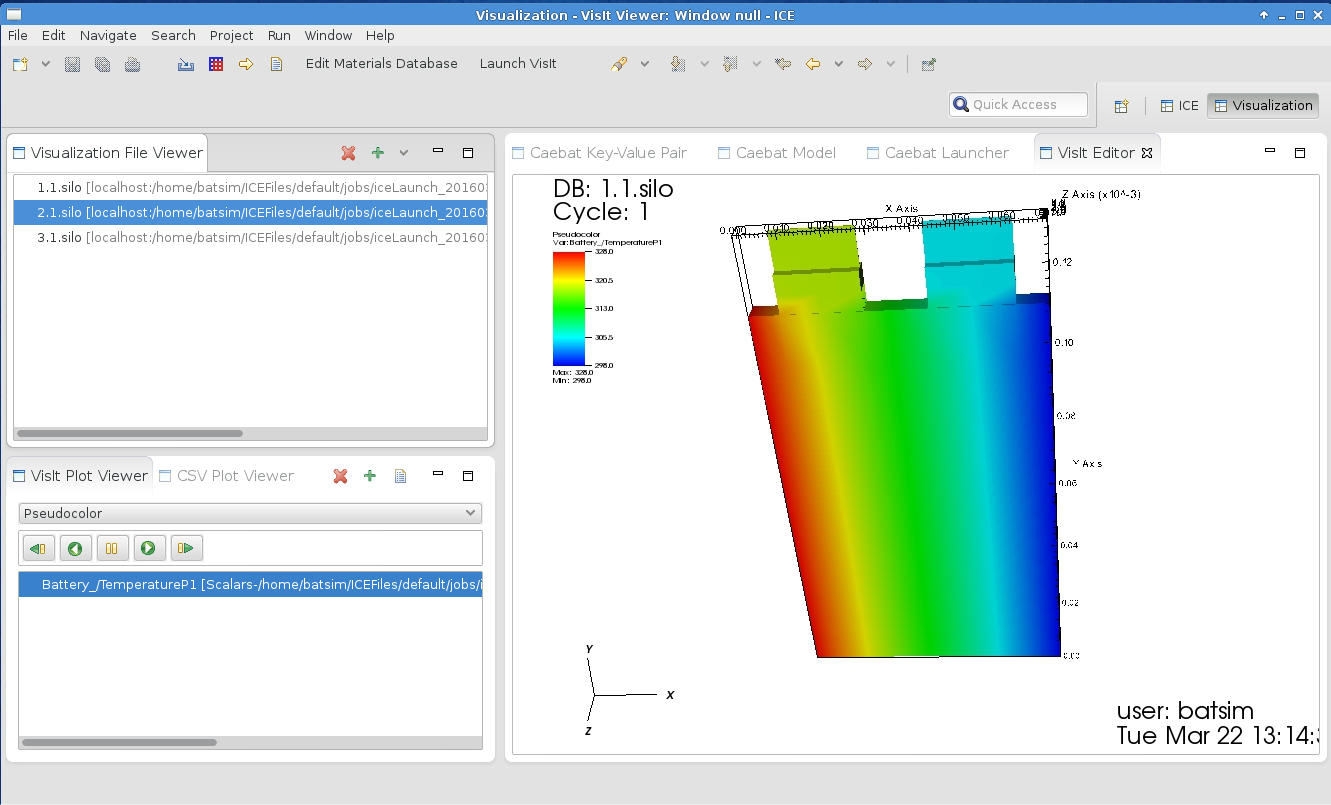}
\caption{ICE workbench for VIBE analysis.}
\end{figure}

VIBE 1.0 is available as a virtual machine (for convenient deployment) in which the simulation
software and ICE are provided side by side. The VIBE team's more recent 
efforts for VIBE~1.1 include providing the simulation
software in Docker containers so that users can download the latest,
native version of ICE for their machine while simultaneously benefiting
from a smaller virtual machine for the simulator.

\subsection{Multiphysics Simulations with
MOOSE}\label{multiphysics-simulations-with-moose}

The MOOSE Framework is a powerful, easy-to-use multiphysics framework
developed at Idaho National Laboratory \cite{gaston_moose:_2009}. ICE
provides workflow tools for MOOSE as well as specialized class
generation utilities for developing custom MOOSE kernels. Many of the
MOOSE tools in ICE were developed closely with the MOOSE team to
reproduce various aspects of MOOSE's user interface, known as ``Peacock.'' 
Figure 2 shows an example of the ICE workbench for a simple structural mechanics
problem solved using the MOOSE framework \cite{mccaskey_scientific_2015}.

\begin{figure}[htbp]
\centering
\includegraphics[width=\textwidth]{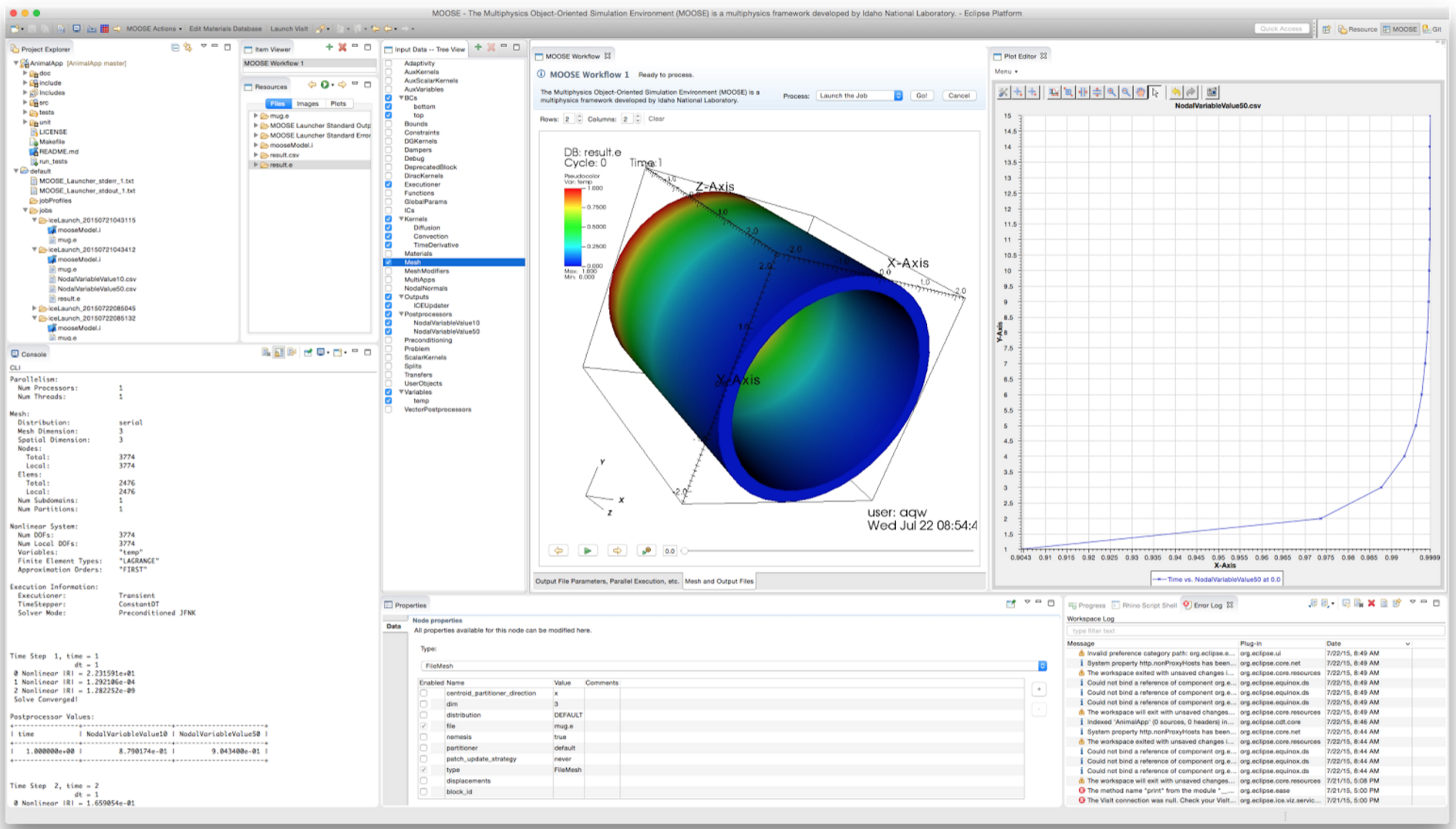}
\caption{ICE workbench for MOOSE workflows.}
\end{figure}

There are over three hundred MOOSE-based applications, and it is very
easy to create new MOOSE applications. The ICE development team uses ICE
and MOOSE to quickly solve energy science problems with HPC resources 
and to deploy domain-specific workbenches. ICE provides features for automatic installation, configuration and optimization of scientific development environments. In the context of MOOSE, ICE includes support for automatically downloading and building MOOSE from MOOSE's GitHub
repository. This integration enables users to immediately begin developing
complex multiphysics applications using the pre-installed Eclipse C Development 
Tools. Once the new MOOSE-based application is built, it will automatically work with
the MOOSE workflow tools in ICE, although developers can also create
customized workflow tools as needed.

\subsection{Binder Jet Modeling}\label{binder-jet-modeling}

Solid-state sintering of parts printed using binder jetting
significantly increases part strength by decreasing the part porosity
and eliminating voids. However, this process does cause the resulting
product to shrink and warp from its original layout. An ideal, near-net-shape, 
process combines binder jetting with solid-state sintering and accounts
for warpage and deformation in the part's design phase. 
Figure 3 shows an ICE-based workbench for performing simulations of
this process with visualizations of the pre- and post-simulation
properties of a central body with eight cantilevers. The primary
deformation in this type of geometry is bending or drooping of the
cantilevers, and ICE currently calls a custom MOOSE-based application
(which was written in ICE as described above) for simulating the
deformation of the cantilevers. The full application for this work 
is expected to be released near the end of 2017.

\begin{figure}[H]
\centering
\includegraphics[width=\textwidth]{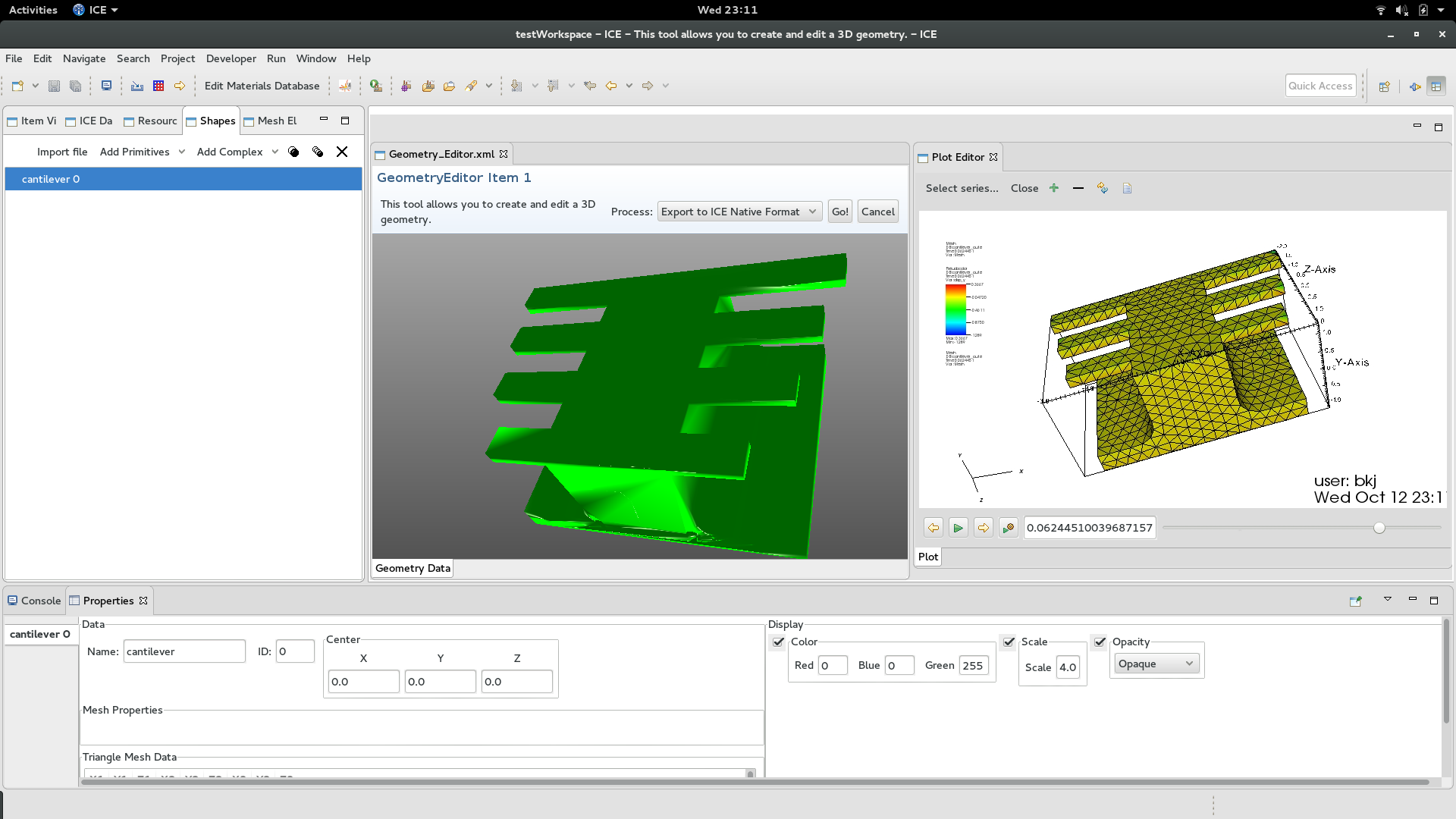}
\caption{ICE workbench for binder jet modeling.}
\end{figure}

\subsection{Neutron Reflectivity}\label{neutron-reflectivity}

ICE also includes a small utility for simulating neutron reflectivity and
comparing the results to other data \cite{billings_brand_2015}. This 
utility was developed in collaboration with a team at ORNL's Spallation Neutron 
Source to replace an older utility that was originally written in Visual Basic and 
distributed via Excel macros. The new utility, developed in ICE and 
shown in Figure 4, is capable of processing a single workflow.

\begin{figure}[H]
\centering
\includegraphics[width=\textwidth]{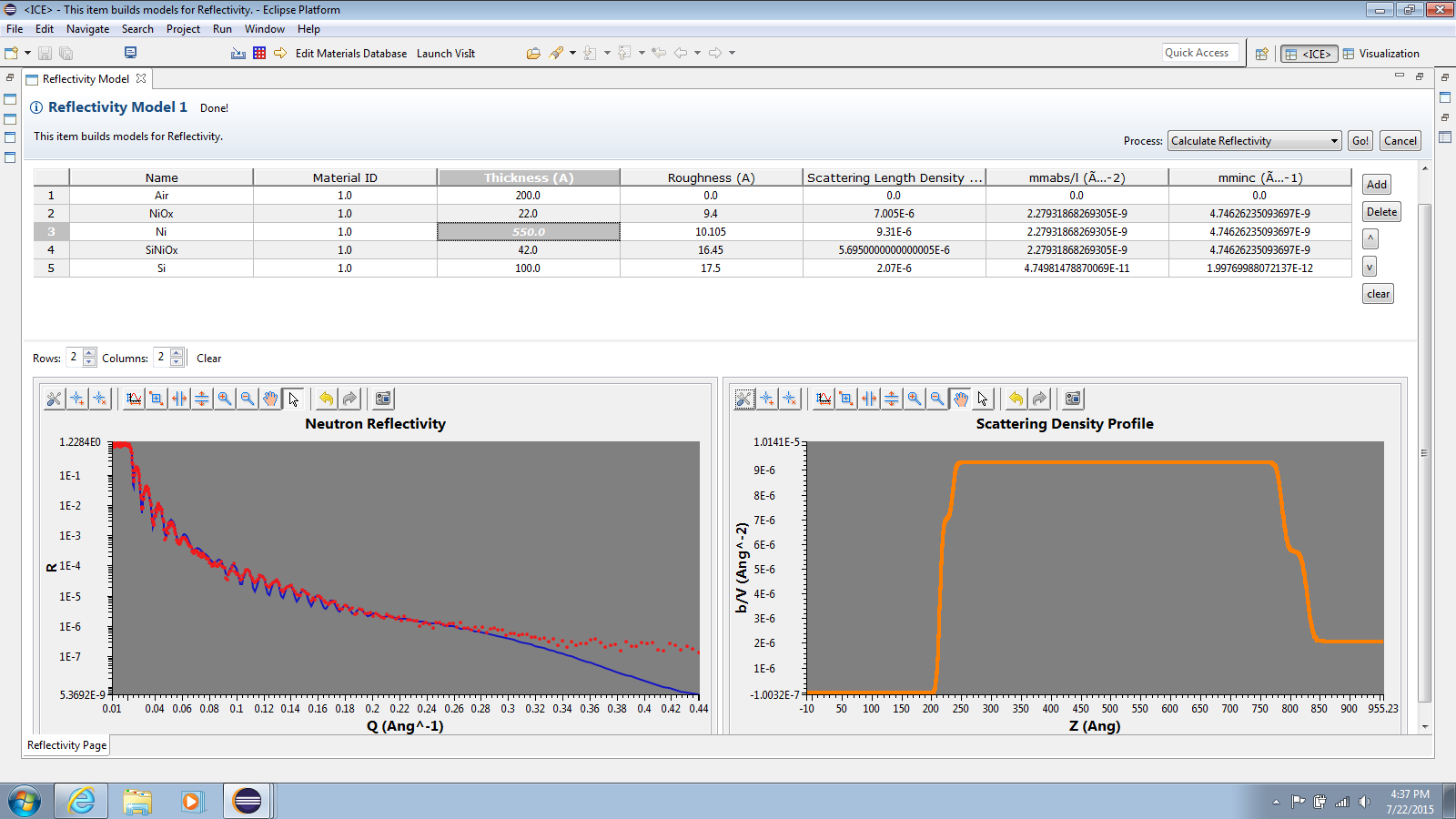}
\caption{ICE workbench for neutron reflectivity.}
\end{figure}

\subsection{Quantum Computing}\label{quantum-computing}

As quantum computing grows, the need for sophisticated
software that can utilize quantum hardware or perform calculations on
simulated quantum hardware becomes more pressing. Humble et al. created a simulator for
adiabatic quantum computers where workflows were added to ICE to support
interaction with the simulator and to process large sets of
quadratic binary optimization problems \cite{humble_integrated_2014}. Figure 5
shows the workbench for this project.

\begin{figure}[H]
\centering
\includegraphics[width=\textwidth]{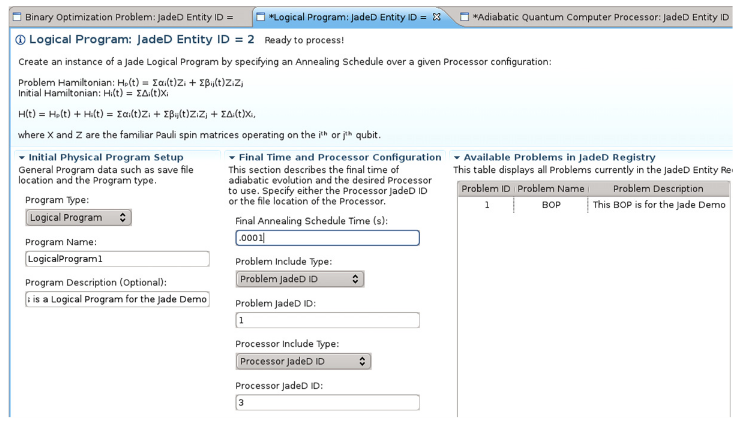}
\caption{ICE workbench in the Jade Adiabatic Development Environment (JADE) for quantum computing simulations.}
\end{figure}

ICE also supports other quantum computing projects, including the
eXtreme-scale ACCelerator (XACC) effort
\cite{mccaskey_ornl-qci/xacc_2016}. XACC is a programming framework for
extreme-scale, post-exascale accelerator architectures that can be
integrated alongside existing, conventional applications.

\subsection{Nuclear Energy}\label{nuclear-energy}

There are many examples of ICE's role in modeling and simulation
projects for nuclear energy, but for an example of the level of
customization that is possible in ICE, readers are referred to the work outlined in
\cite{billings_domain-specific_2015}. Support for the ``Reactor
Analyzer'' was dropped in ICE 2.1.8, but it demonstrated ICE's ability
to integrate many different nuclear energy tools for complex analysis.

\section{Impact}\label{impact}

The impact of software tools like ICE is difficult to quantify. However,
there are several examples of when ICE has significantly assisted its
development team and others.

One pressing area of interest and impact is that of interoperability
between workflow systems. There have been significant prior efforts in combining
large workflow management systems (e.g., Mandal et al.
\cite{mandal_integrating_2007}), but the end goal of gaining the greatest
advantage by using the best capabilities from multiple systems remains elusive.
ICE's unique perspective on workflows and its well-defined API make it
possible to integrate multiple systems in a straightforward way. This
allows it to connect to other workflow environments, such as Triquetrum,
quite easily \cite{brooks_introducing_2016}. Triquetrum, like Kepler in
\cite{mandal_integrating_2007}, is a Ptolemy-based workflow engine
\cite{brooks_triquetrum:_2015}. Several of the authors of this paper are using
it to investigate these issues.

It is widely known that tools that enable researchers to be more productive tend
to improve the pursuit of existing research. The high extensibility of
ICE and the tools that it combines from the larger Eclipse ecosystem
have made it possible for researchers on the development team to quickly
deploy new simulation environments for their research problems (e.g., the binder jet
modeler and reflectivity tools mentioned earlier). Other tools created
with ICE may not invent something radically new, but they tend to
streamline interactions with those tools. Many ICE users, and certainly
the development team, have experienced improvements in their software
development efforts because of the tools that ICE provides or have learned
new technologies because access to new tooling was as simple as
installing more plugins through the Eclipse Marketplace.

The ICE development team does not track ICE's user base, as useful as
that would be, because of the extra work involved. However, various
sources such as the VIBE mailing list, ICE's own mailing lists, and
website download statistics suggest that ICE has been used by over 350
people at one time or another and currently has about twenty
``superusers,'' including the development team.

A new cloud-based development tool based on ICE is under development by
RNET Technologies, Inc. out of Dayton, Ohio in response to a Small Business Innovation 
Research award from the US Department of Energy. This web-based version of 
ICE will continue ICE's support for nuclear energy and will integrate with cloud 
computing solutions like Amazon Web Services and ORNL's Compute and Data
Environment for Science (CADES). Additionally, while ICE has not directly led to
any ``spin-off'' companies, ICE source code has been used
in two spin-off projects: (1) EAVP (mentioned earlier) for advanced
visualizations and (2) the Eclipse January project for scientific data
structures \cite{graham_eclipse_2016}. ICE was also one of the founding
projects of the Eclipse Science Working Group.

Eclipse ICE is an open-source project, and the authors welcome and encourage
engagement and contributions from users. Interested parties may inquire by
contacting the corresponding author or visiting the ICE website.

\section{Sample Code, Tutorials, and Other
Resources}\label{sample-code-tutorials-and-other-resources}

The primary resource for information on ICE is the project's 
website~\cite{billings_eclipse_2016}. The ``Resources'' menu includes links to
detailed tutorials and user documentation. Examples of how to create
new workflow Items are available at
\url{https://github.com/eclipse/ice/tree/master/org.eclipse.ice.demo}.
Examples of how to use the scripting engine are available at
\url{https://github.com/eclipse/ice/tree/master/examples}. ICE also includes an
extensive suite of unit, integration, and UI tests, which are
also excellent examples of how to work with the platform. Tutorial and demonstration videos are available on YouTube at \url{https://goo.gl/nxCzRD}.

\section{Conclusions}\label{conclusions}

Modeling and simulation workflows for computational scientists differ
greatly from those of experimentalists or those who primarily interact with
grid-based workflow management systems. The Eclipse ICE effort described here 
has been used to address interdisciplinary problems in 
modeling and simulation for energy science. Of particular interest are the
differences in architecture between a workflow management system focused
on modeling and simulation compared to those focused on grids. ICE's broad
applicability across many topics in energy science suggests that
there are opportunities for these systems in general. Finally, one 
interesting avenue of future exploration is coupling or integrating ICE 
with other workflow tools such as Aiida, Triquetrum, Kepler, and Pegasus,
which would make it possible to combine the best of grid and modeling workflows and
simulation workflows to address greater challenges.

\section*{Acknowledgments}\label{acknowledgments}
\addcontentsline{toc}{section}{Acknowledgments}

The authors are grateful for the assistance and support of the following
people and institutions without whom this work would not have been
possible. This includes many people who directly contributed to the
project, either in its early days as ``NiCE'' or once it moved to
Eclipse, including: Ronald Allen, Andrew Belt, Tim Bohn, David E.
Bernholdt, Erica Grant, Mike Guidry, Forest Hull, Eric J. Lingerfelt,
Sebastien Jourdain, JiSoo Kim, Allison Koenecke, Fangzhou Lin, Greg
Lyon, Tony McCrary, John M. Hetrick III, Elizabeth Piersall, Neeti
Pokhriyal, Adrian Sanchez, Claire Saunders, Nick Stanish, Matthew Wang, and
Scott Wittenberg. The support of both the Nucleare Energy Advanced Modeling and Simulation (NEAMS) and Consortium for Simulation of Light-water Reactors (CASL) programs is also
greatly appreciated. The authors would like to acknowledge the special
contribution of the Eclipse Foundation, the Eclipse Community, the
Eclipse Science Working Group, and our many colleagues who use and
contribute to open-source projects in the Eclipse ecosystem.

Finally, the development team is especially grateful to Barney Maccabe,
David Pointer, and John Turner of ORNL for their
endless support and advocacy for this work.

This work has been supported by the US Department of Energy, Offices of
Nuclear Energy (DOE-NE) and Energy Efficiency and Renewable Energy
(DOE-EERE), and by the ORNL Undergraduate Research Participation
Program, which is sponsored by ORNL and administered jointly by ORNL and
the Oak Ridge Institute for Science and Education (ORISE). This work was
also supported in part by the ORNL Director's
Research and Development Fund. ORNL is managed by UT-Battelle, LLC, for
the US Department of Energy under contract no. DE-AC05-00OR22725. ORISE
is managed by Oak Ridge Associated Universities for the US Department of
Energy under contract no. DE-AC05-00OR22750.

\section*{Required Metadata}\label{required-metadata}
\addcontentsline{toc}{section}{Required Metadata}

\section*{Current Code Version}\label{current-code-version}
\addcontentsline{toc}{section}{Current code version}

See Table \ref{codeTable}.

\begin{table}[H]
\begin{tabularx}{\textwidth}{|l|X|X|}
\hline
C1 & Current code version & `next'\tabularnewline\hline
C2 & Permanent link to code/repository used for this code version &
\url{https://github.com/eclipse/ice/tree/next}
\tabularnewline\hline
C3 & Legal code license & Eclipse public license 1.0 \tabularnewline\hline
C4 & Code versioning system used & Git \tabularnewline\hline
C5 & Software code languages, tools, and services used & Java, OSGi, Eclipse RCP,
and Maven \tabularnewline\hline
C6 & Compilation requirements, operating environments and dependencies & Java 1.8 or greater, Maven, and
an internet connection for dependencies \tabularnewline\hline 
C7 & If available, link to developer documentation/manual &
\url{https://wiki.eclipse.org/ICE} \tabularnewline\hline 
C8 & Support email for questions & ice-dev@eclipse.org \tabularnewline\hline
\end{tabularx}
\caption{Current Code Version}
\label{codeTable}
\end{table}

\section*{Current Executable Software
Version}\label{current-executable-software-version}
\addcontentsline{toc}{section}{Current Executable Software Version}

See Table \ref{execTable}.

\begin{table}[H]
\begin{tabularx}{\textwidth}{|l|X|X|}
\hline
S1 & Current code version & 2.2.1 \tabularnewline\hline
S2 & Permanent link to executables of this version &
\url{https://www.eclipse.org/downloads/download.php?file=/ice/builds/2.2.1/}
 \tabularnewline\hline 
S3 & Legal software license & Eclipse public license 1.0 \tabularnewline\hline
S4 & Computing platforms/operating systems & Windows (32/64-bit), Mac OS/X,
Linux (32/64-bit) \tabularnewline\hline 
S5 & Installation requirements and dependencies & Java 1.8 or
greater \tabularnewline\hline
S6 & If available, link to user manual. If formally published, include a
reference to the publication in the reference list &
\url{https://wiki.eclipse.org/ICE} \tabularnewline\hline 
S7 & Support email for questions & ice-users@eclipse.org \tabularnewline\hline
\end{tabularx}
\caption{Current Executable Software}
\label{execTable}
\end{table}